\documentclass[preprint,11pt,5p,times,twocolumn,authoryear]{elsarticle}
\usepackage{cuted}
\usepackage{amssymb}
\usepackage{lipsum}
\usepackage{cuted}
\usepackage{amsmath}
\usepackage{xcolor}  
\usepackage[
    colorlinks=true,
    citecolor=blue,   
    linkcolor=red,    
    urlcolor=magenta 
]{hyperref} 
\usepackage{booktabs}

\usepackage{hyperref}

\journal{Journal of High Energy Astrophysics}

\begin{document}

\begin{frontmatter}

%% Title, authors and addresses

%% use the tnoteref command within \title for footnotes;
%% use the tnotetext command for theassociated footnote;
%% use the fnref command within \author or \affiliation for footnotes;
%% use the fntext command for theassociated footnote;
%% use the corref command within \author for corresponding author footnotes;
%% use the cortext command for theassociated footnote;
%% use the ead command for the email address,
%% and the form \ead[url] for the home page:
%% \title{Title\tnoteref{label1}}
%% \tnotetext[label1]{}
%% \author{Name\corref{cor1}\fnref{label2}}
%% \ead{email address}
%% \ead[url]{home page}
%% \fntext[label2]{}
%% \cortext[cor1]{}
%% \affiliation{organization={},
%%            addressline={}, 
%%            city={},
%%            postcode={}, 
%%            state={},
%%            country={}}
%% \fntext[label3]{}

\title{Investigating the Temporal Evolution of Gamma-Ray Burst Central Engine Parameters Based on Numerical Simulations} %% Article title

%\title{Hyperaccreting Black Hole as Gamma-Ray Burst Central Engine: the Numerical Simulation and Analytical Results} %% Article title

%% use optional labels to link authors explicitly to addresses:
%% \author[label1,label2]{}
%% \affiliation[label1]{organization={},
%%             addressline={},
%%             city={},
%%             postcode={},
%%             state={},
%%             country={}}
%%
%% \affiliation[label2]{organization={},
%%             addressline={},
%%             city={},
%%             postcode={},
%%             state={},
%%             country={}}

\author{Wei-Hua Lei} %% Author name
\ead{leiwh@hust.edu.cn}
% Corresponding author text
%\cortext[1]{Corresponding author}

% Footnote text
\fntext[1]{Preprint submitted to Journal of High Energy Astrophysics}

%% Author affiliation
\affiliation{organization={Department of Astronomy, School of Physics, Huazhong University of Science and Technology},%Department and Organization
            %addressline={}, 
            city={Wuhan},
            postcode={430074}, 
            %state={},
            country={China}}

%% Abstract
\begin{abstract}
A hyperaccreting stellar-mass black hole (BH) has been proposed as the candidate central engine of gamma-ray bursts (GRBs). Comparing the predictions from the central engine models with the temporal behavior of GRBs is of great interest. In this paper, using the open-source GRMHD HARM-COOL code, we evolve several 2D magnetized hyperaccreting BH models with realistic equation of state in a fixed curved space-time background. We extend the code to include the calculation of neutrino annihilation power. We then study the time evolution of BH central engine parameters, i.e., the neutrino annihilation power, the Blandford-Znajke (BZ) power, and the initial magnetization $\sigma_0$. We find that the neutrino power is generally consistent with previous analytical results. Usually, the neutrino annihilation process tends to launch a thermal ``fireball'', while the BZ jet is Poynting-flux-dominated. Our results, especially the evolution characteristics of $\sigma_0$ may help to understand the complex GRB spectral behavior.
\end{abstract}

%%Graphical abstract
%\begin{graphicalabstract}
%\includegraphics{figs/grabs.pdf}
%\end{graphicalabstract}

%%Research highlights
%\begin{highlights}
%\item Research highlight 1
%\item Research highlight 2
%\end{highlights}

%% Keywords
\begin{keyword}
%% keywords here, in the form: keyword \sep keyword

%% PACS codes here, in the form: \PACS code \sep code

%% MSC codes here, in the form: \MSC code \sep code
%% or \MSC[2008] code \sep code (2000 is the default)
gamma-ray bursts \sep black hole \sep accretion disk 
\end{keyword}

\end{frontmatter}

%% Add \usepackage{lineno} before \begin{document} and uncomment 
%% following line to enable line numbers
%% \linenumbers

%% main text
%%

%% Use \section commands to start a section
\section{Introduction}
\label{sec:introduction}
%% Labels are used to cross-reference an item using \ref command.
Rich observations have greatly enhanced our knowledge of gamma-ray bursts (GRBs). The association between long GRBs and Type Ic supernovae suggests their origin of massive star core-collapse \citep{Woosley1993,Paczynski1998,MacFadyen1999}, and the occurrence of GW170817 accompanied by GRB 170817A \citep{Abbott2017,Kasliwal2017} supports the binary merger (neutron star-neutron star, or neutron star-black hole) origin model of short GRBs invoking mergers of two neutron stars (NSs) or of a black hole (BH) and an NS \citep{Eichler1989,Paczynski1991,Fryer1999,Li2016}. However, the nature and property of GRB central engines are still poorly understood.

%Gamma-ray bursts (GRBs) are The nature and properties of gamma-ray burst (GRB) central engines  association between long Gamma-ray bursts (LGRBs) 

%Long GRBs are connected with core-collapse supernovae

%The  Gamma-ray bursts (GRBs) are widely. It is widely believed that long GRBs are connected with core-collapse supernovae, and short GRBs are likely related to mergers of two neutron stars (NSs) or of an NS and a black hole (BH). These scenarios lead to the formation of a stellar-mass BH or a millisecond magnetar.

%The complex nature of macroscopic and microphysical properties of the central engines requires a still ongoing effort to identify crucial aspects of their operation.

%Two types of GRB central engine models have beendiscussed in the literature, i.e., the BH model and magnetar model. 
The popular GRB central engine model invokes a stellar-mass BH surrounded by a hyperaccretion disk (the so-called neutrino-cooling-dominated accretion flow, NDAF). The relativistic jet in a GRB can be powered by the neutrino–antineutrino annihilation mechanism \citep{PWF99,DPN02,Gu2006,Chen2007,Janiuk2007,Lei2009,Cao2014,Xie2016,Liu2007,Liu2017}, or the Blandford–Znajek \citep[hereafter BZ]{BZ77}
mechanism \citep{Lee2000,Li2000,Lei2005,Lei2011,Lei2013,Lei2017,Liu2018}.
It is crucial to differentiate these different mechanisms using observational data. 

The temporal behavior of GRBs in the prompt emission phase may provide
meaningful clues to the central engine models. In some GRBs (e.g., GRB 080916C), the spectra studies show no evidence of quasi-thermal emission \citep{Abdo2009}. This motivates us to investigate the evolution of the BH central engine and compare their predictions with the temporal behavior of GRBs.  

In our previous work, we addressed the fundamental problem of baryon loading in GRB jets, and found that a magnetically dominated jet can be much cleaner and is more consistent with the requirement of large Lorentz factors in GRBs \citep{Lei2013}. We then investigated the BH central engine parameters, such as the jet power, the dimensionless entropy $\eta$, and the central engine parameter $\mu_0=\eta (1+\sigma_0)$ (where $\sigma_0$ is the initial magnetization of the engine) at the base of the jet \citep{Lei2017}. With these results, a number of works adopted some empirical correlations (such as jet power vs. Lorentz factor $\Gamma_0$ and minimum variability timescale vs. the Lorentz factor $\Gamma_0$)  \citep{Xie2017,Yi2017,Yi2021,Nicole2018,Li2018}, or the spectral evolution characteristics of prompt emissions \citep{Gao2022,ZhangB2024,Fu2024,Li2024}, to differentiate the central engine models. However, these studies are mainly based on analytical solutions to the NADF and BZ models.

The general relativistic magnetohydrodynamic (GRMHD) code HARM (High Accuracy Magnetohydrodynamics) is a finite-volume code with HLL shock capturing scheme, which works on the stationary metric around the BH. The code integrates the total energy equation and updates the set of ``conserved'' variable, i.e., comoving density, energy-momentum, and magnetic field \citep{Gammie2003}. \citet{Janiuk2017} self-consistently incorporated the detailed nuclear equation of state (the degeneracy of relativistic electrons, protons,
and neutrons) into the HARM scheme \citep{Janiuk2007,Janiuk2017,Janiuk2019}. With this developed code, namely HARM-COOL, they investigated the temporal evolution of the neutrino flux, the BZ power and r-process nucleosynthesis in the outflow \citep{Janiuk2017,Janiuk2019,Sapountzis2019}. However, the estimation of the neutrino annihilation power, which is fundamental in the study of the central engine of GRBs, is lacking in the code.

In this paper, we perform 2D GRMHD simulations for the hyperaccreting BH with HARM-COOL, and include the treatment of neutrino annihilation. We then investigate the evolution of the central engine parameters, such as the jet power and the initial magnetization.  

This paper is organized as follows. In Section 2, we present the simulation scheme and the initial configuration of the HARM-COOL. The calculation of neutrino annihilation luminosity, BZ power and magnetization parameter $\sigma_0$ are given in Sections 2 and 3. Finally, we summarize our results and discuss some related issues in Section 4.

\section{The Simulation Setup}
\label{sec:setup}
The GRMHD code HARM provides a solver for continuity and energy-momentum conservation equations \citep{Gammie2003,Noble2006}
\begin{eqnarray}
    (\rho u^\mu)_{;\mu}=0, \nonumber \\
    T^\mu_{\nu;\mu}=0,
\end{eqnarray}
and induction equation,
\begin{eqnarray}
    \partial_t(\sqrt{-g}B^i) = -\partial_j[\sqrt{-g}(b^j u^i - b^i u^j)],
\end{eqnarray}
where $T^{\mu\nu}=T^{\mu\nu}_{\rm gas}+T^{\mu\nu}_{\rm EM}$, $T^{\mu\nu}_{\rm gas}=(\rho+u+p)u^\mu u^\nu+pg^{\mu\nu}$, $T^{\mu\nu}_{\rm EM}=b^2 u^\mu u^\nu+b^2 g^{\mu\nu}/2-b^\mu b^\nu$, and $b^\mu=u_\nu {^*F^{\mu\nu}}$. Here, $u^\mu$ is the four-velocity of gas, $u$ is the internal energy density, $b^\mu=\epsilon^{\mu\nu\rho\sigma}u_\nu F_{\rho\sigma}$ is the magnetic four-vector, and $F$ is the electromagnetic stress tensor. Assuming the fore-free approximation, we have $E_\nu=u_\mu F^{\mu\nu}=0$.

%$T^{\mu\nu}_{\rm gas}=\rho h u^\mu u^\nu+pg^{\mu\nu}=(\rho+u+p)u^\mu u^\nu+pg^{\mu\nu}$,

The conservative numerical scheme used in this code solves $\partial_t \mathbf{U(P)}=-\partial_i \mathbf{F}^i\mathbf{(P)+S(P)}$, where $\mathbf{U}$ is a vector of ``conserved'' variables (momentum, energy density, number density, taken in the coordinate frame), $\mathbf{P}$ is a vector of ``primitive'' variables (rest-mass density, internal energy), $\mathbf{F}^i$ are the fluxes, and $\mathbf{S}$ is a vector of source terms. Inversion $\mathbf{P(U)}$ is calculated, therefore, in every time step, numerically.
%The HARM scheme solves for the inversion between the ``primitive'' and ``conserved'' variables at every time step.

In HARM-COOL, the nuclear equation of state is incorporated into the code \citep{Janiuk2017,Janiuk2019}. The total pressure is contributed by free nucleons, pairs, radiation, alpha particles, and also trapped neutrinos \citep{Yuan2005,Janiuk2007,Liu2017}, and is given by:
\begin{eqnarray}
    P_{\rm gas} = P_{\rm nucl}+ P_{\rm He}+P_{\rm rad} +P_{\nu},
\end{eqnarray}
where $P_{\rm nucl} = P_{\rm e^-}+P_{\rm e^+}+P_{\rm n} +P_{\rm p}$ and
\begin{eqnarray}
    P_{i} = \frac{2\sqrt{2}}{3\pi^2}\frac{(m_{i} c^2)^4}{(\hbar c)^3} \beta_i^{5/2}[F_{3/2}(\eta_i,\beta_i) +\frac{1}{2} \beta_i F_{5/2}(\eta_i,\beta_i)],
\end{eqnarray}
where $F_k$ are the Fermi-Dirac integrals of the order $k$,  and $\beta_i=kT/m_i c^2$ are the relativity parameters. $\eta_{\rm e}$, $\eta_{\rm p}$, and $\eta_{\rm n}$ are the reduced chemical potentials of electrons, protons, and neutrons, respectively (where $\eta_i=\mu_i/kT$, also known as the degeneracy parameter, $\mu_i$ the standard chemical potential). $\eta_{\rm e^+}=-\eta_{\rm e}-2/\beta_{\rm e}$ is the reduced chemical potential of positrons. The pressure of helium is taken as ideal gas, $P_{\rm He} = (n_{\rm b}/4)(1-X_{\rm nuc})kT$, where $X_{\rm nuc}$ is the mass fraction of free nucleons. The radiation pressure, $P_{\rm rad}=(4\sigma)/(3c)T^4$. When neutrinos become trapped in the disk, the neutrino pressure $P_\nu$ is nonzero \citep{DPN02,Liu2017},
\begin{equation}
P_\nu = \frac{7\sigma T^4}{6c}\sum {\frac{\tau _{\nu _i } / 2 + 1 /
\sqrt 3}{\tau _{\nu _i } / 2 + 1 /
\sqrt 3 + 1 / (3\tau _{a,\nu _i } )}} ,
\label{eq:Pv}
\end{equation}
where $\tau_{\nu _i }=\tau_{{\rm a},{\nu _i }}+ \tau_{\rm s}$, $\tau_{{\rm a},{\nu _i }}$ and $\tau_{\rm s}$ denote absorption and scattering optical depth. Here $\tau_{{\rm a},{\nu _i }}=(H/((7/8)\sigma T^4)) q_{{\rm a},{\nu_i}}$, where $q_{{\rm a},{\nu_e}}=q_{\rm pair}+q_{\rm urca}+q_{\rm plasm}+q_{\rm brems}/3$ and $q_{{\rm a},{\nu_\mu}}=q_{\rm pair}+q_{\rm brems}/3$. The scattering optical depth is given by $\tau_{\rm s}=24.3\times 10^{-5} (kT/m_{\rm e} c^2)^2 H (C_{\rm s,p} n_{\rm p} +C_{\rm s,n} n_{\rm n})$, with $C_{\rm s,p}=[4(C_V-1)^2+5\alpha^2]/24$, $C_{\rm s,n}=(1+5\alpha^2)/24$, $C_V=1/2+2\sin^2\theta_C$, $\alpha=1.25$ and $\sin^2\theta_C=0.23$.

The cooling rates due to the bremsstrahlung
and plasmon decay, are \citep{DPN02}
\begin{equation}
q_{\rm brems} = 3.35\times 10^{27}\rho^2_{10} T^{5.5}_{11} ,
\label{eq:qbrems}
\end{equation}
\begin{equation}
q_{\rm plasm} = 1.5\times 10^{32} T^{9}_{11} \gamma^6_{\rm p} e^{-\gamma_{\rm p}} (1+\gamma_{\rm p}) \left(2+ \frac{\gamma^2_{\rm p}}{1+\gamma_{\rm p}} \right),
\label{eq:qbrems}
\end{equation}
where $\gamma_{\rm p}=5.565\times 10^{-2} [[\pi^2+3(\mu_e/kT)]/3]^{1/2}$. The cooling rates due to URCA processes, $q_{\rm urca}$, and pair annihillation, $q_{\rm pair}$, reactions have more complex forms, and
can be found, i.e., in \citet{Janiuk2007}. The disk height $H$ is given by \citep{Janiuk2017},
\begin{equation}
    H=\frac{c}{\Omega_{\rm D}} \sqrt{\frac{P}{\rho c^2+e}},
\end{equation}
where $\Omega_{\rm D} = \frac{c^3}{GM_\bullet}\frac{1}{a_\bullet+(r/r_{\rm g})^{3/2}}$ is the Keplerian frequency, $r_{\rm g}=G M_\bullet/c^2$ and $a_\bullet=J_\bullet c/(G M_\bullet^2)$ is the BH spin parameter. $M_\bullet$ and $J_\bullet$ are the BH mass and angular momentum, respectively. 

%Janiuk et al. (2007; see Equations (A7)–(A15) therein).

The expression for neutrino cooling rate is given by the two-stream approximation $Q_\nu $ is \citep{DPN02,Liu2017}, and it includes the scattering and absorption optical depths for neutrinos of the three flavors,
\begin{equation}
Q_\nu = \sum {\frac{(7 / 8)\sigma T^4}{(3 / 4)(\tau _{\nu _i } / 2 + 1 /
\sqrt 3 + 1 / (3\tau _{a,\nu _i } ))H}} ,
\label{eq:Qv}
\end{equation}
%\textbf{where $\tau_{\nu _i }=\tau_{{\rm a},{\nu _i }}+ \tau_{\rm s}$, $\tau_{{\rm a},{\nu _i }}$ and $\tau_{\rm s}$ denote absorption and scattering optical depth}.
where the total neutrino luminosity of the disk is \citep{Janiuk2017}
\begin{equation}
    \dot{E}_\nu=\int Q_\nu dV,
\end{equation}
where $dV$ is the unit volume in the Kerr geometry.

The detailed discussions over nuclear equation of state and neutrino treatment implemented in HARM-COOL code are given in \citep{Janiuk2013} and \citep{Janiuk2017,Janiuk2019}.

The initial conditions for the accreting material is modeled as an equilibrium torus solution, defined as in \citet[FM]{FM1976}. 
%The initial state of the torus is derived from the analytic solution of the Fisbone-Mocrief (FM) \citep{FM1976} disk model around a Kerr BH. 
The FM disk is defined by $r_{\rm in}$ the inner radius of the disk, $r_{\rm max}$ the radius of maximum pressure, and a dimensionless BH spin $a_\bullet$. 

The BH mass has been set to be $M_\bullet=3M_\odot$ in our simulations. We conveniently adopt the scaling factor, $r_{\rm in}$ and $r_{\rm max}$, so that the FM torus mass of $M_{\rm torus} \sim 0.1 M_\odot$. We seed the torus with a poloidal magnetic field (magnetic field lines follow the constant density surfaces); the strength of the initial magnetic field is normalized by the gas to a magnetic pressure ratio at the pressure maximum of the initial structure of the disk ($\beta\equiv p_{\rm gas}/p_{\rm mag} =50$). We examine three sets of models with different differing with the BH spin ($a_\bullet=0.1,0.6,0.9,0.98$). Table \ref{tb:models} presents a summary of the models and their initial parameters. Our grid resolution is $128 \times 128$.

\begin{table*}[htp]
\begin{center}
\caption{Setup of the Models \label{tb:models}}
\begin{tabular}{ccccccc}
\hline\noalign{\smallskip}
\hline\noalign{\smallskip}
    Model & $a_\bullet$ & $\beta_0$ & $M_\bullet (M_\odot)$ & $M_{\rm t} (M_\odot)$   & $r_{\rm in} (r_{\rm g})$ & $r_{\rm max} (r_{\rm g})$ \\
\hline\noalign{\smallskip}
Mt0.1-a0.1 & 0.1 & 50   & 3 & 0.1 & 7.0 & 14.1  \\
Mt0.1-a0.6 & 0.6 & 50   & 3 & 0.1 & 6.0 & 12.2  \\ 
Mt0.1-a0.9 & 0.9 & 50   & 3 & 0.1 & 5.0 & 10.7  \\
Mt0.11-a0.98 & 0.98 & 50   & 3 & 0.11 & 3.1 & 9.1 \\
\noalign{\smallskip}\hline
\end{tabular}
\end{center}
\end{table*}

\begin{figure*}[ht]%% placement specifier
\centering%% For centre alignment of image.
\includegraphics[width=8cm]{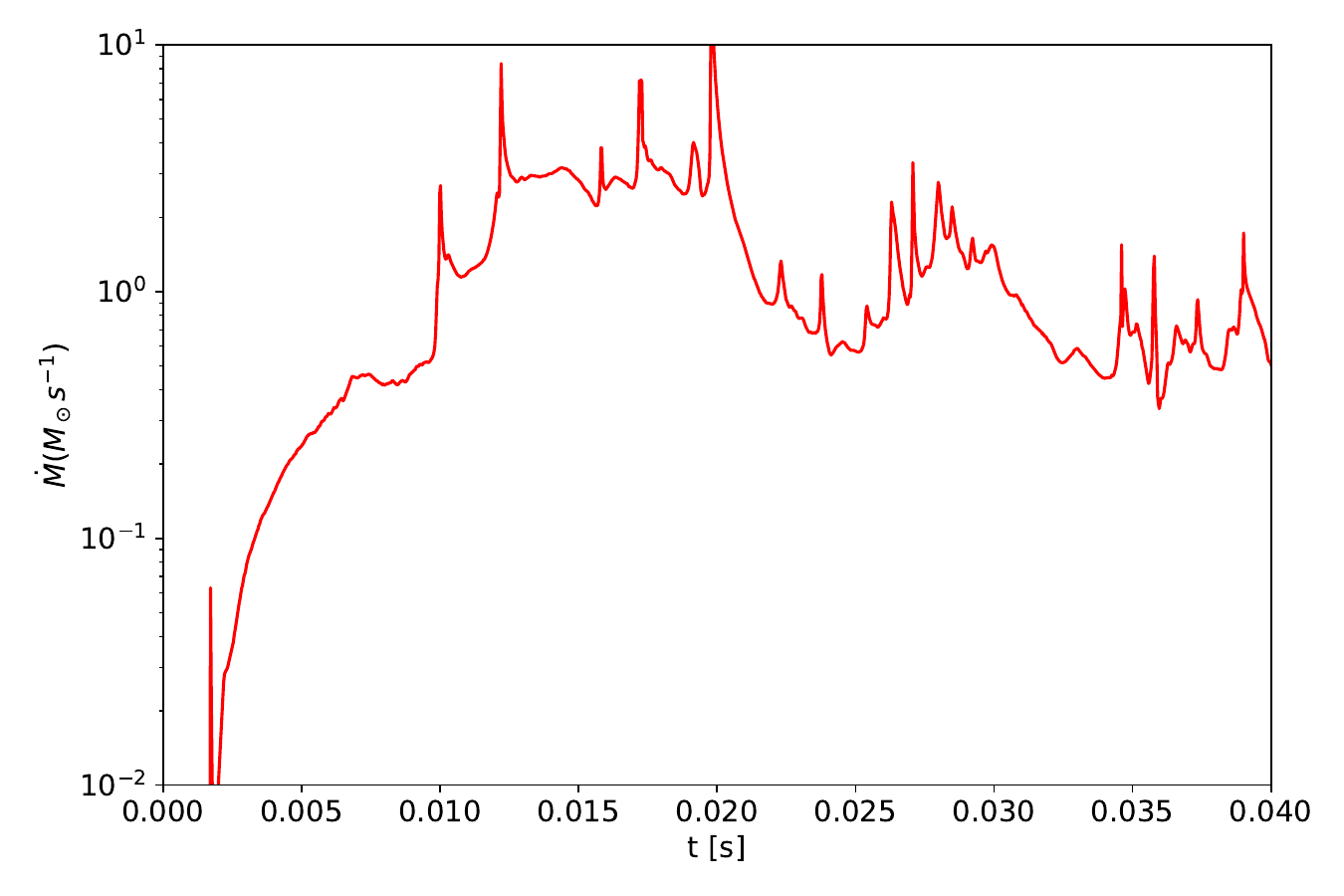}
\includegraphics[width=8cm]{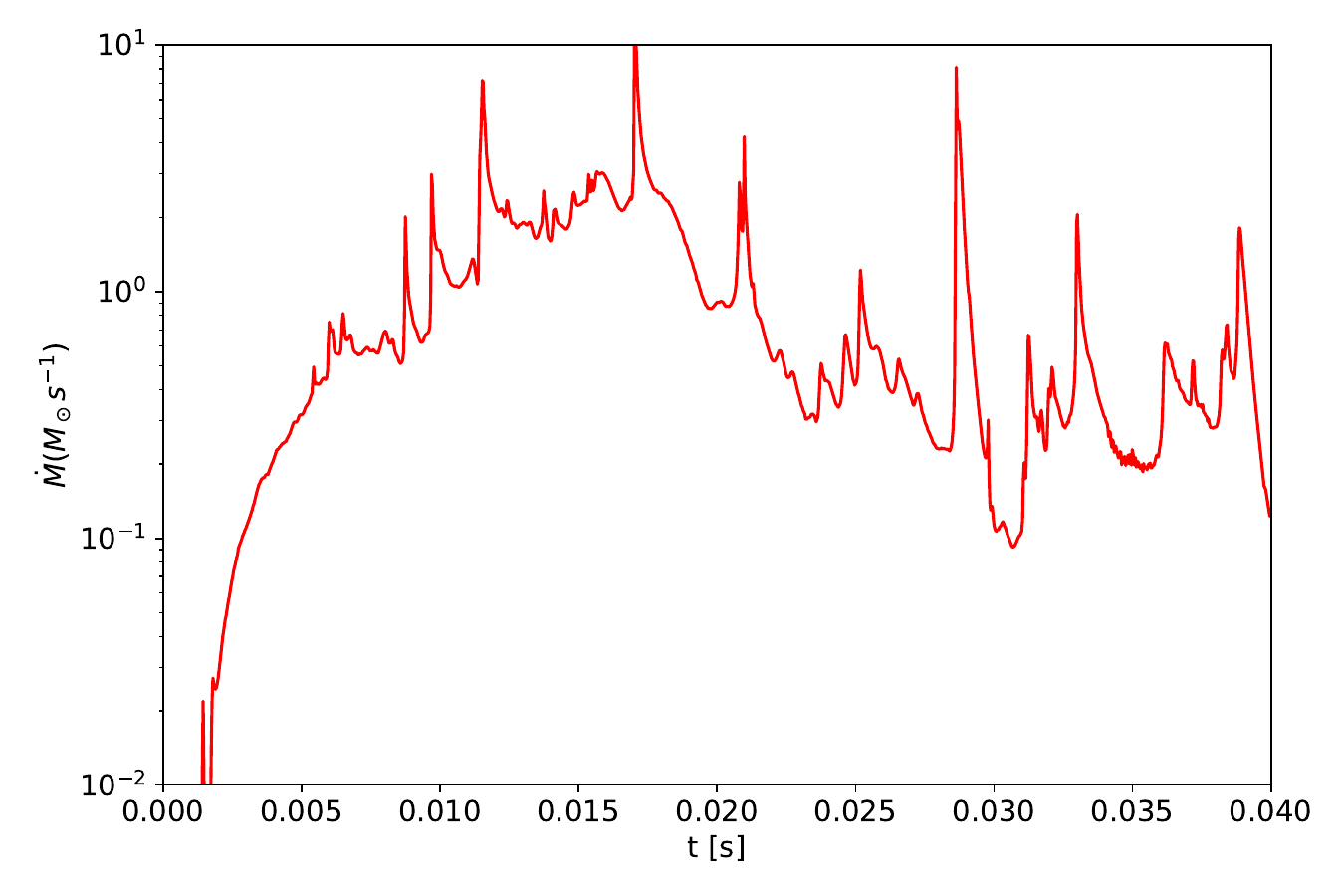}
\includegraphics[width=8cm]{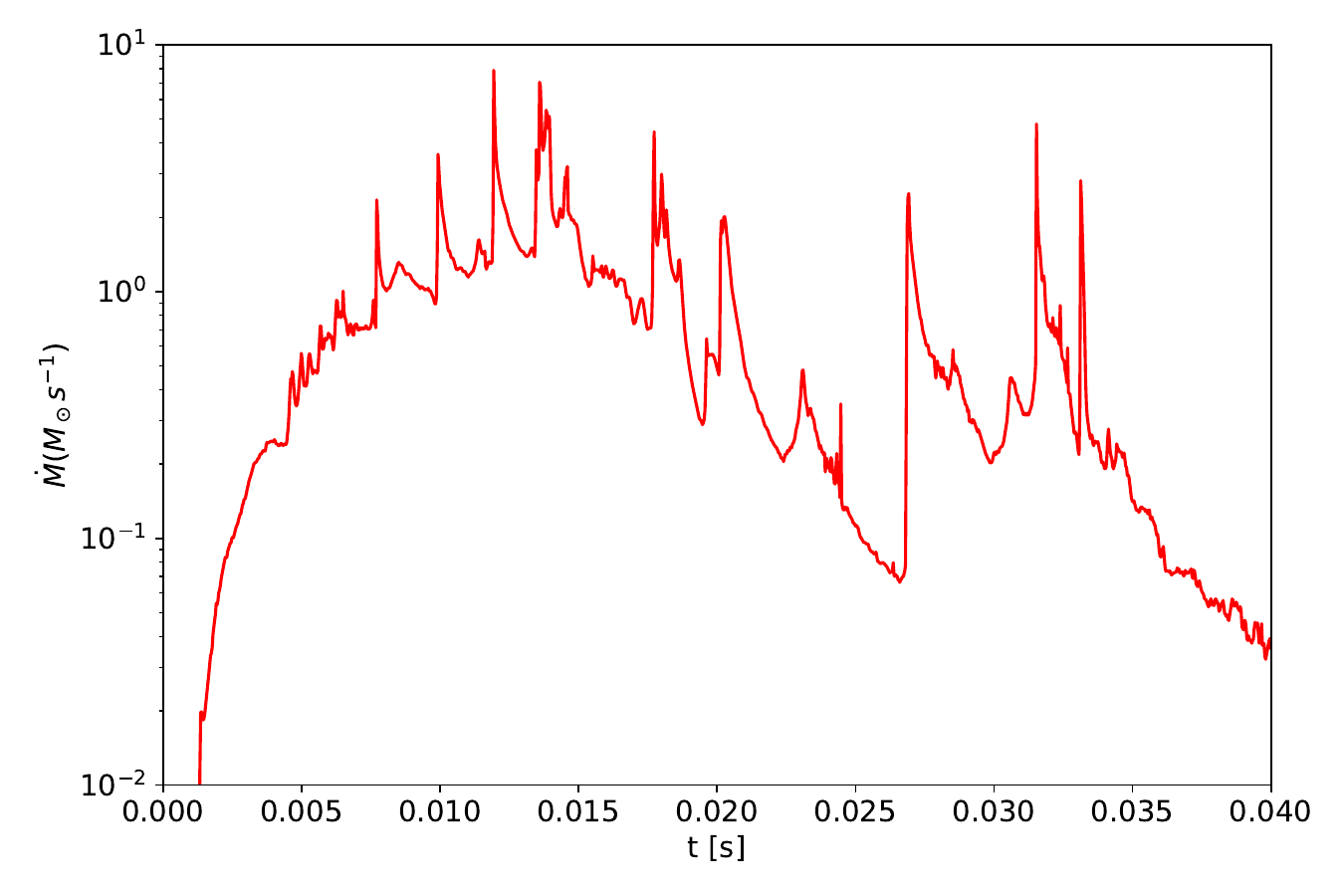}
\includegraphics[width=8cm]{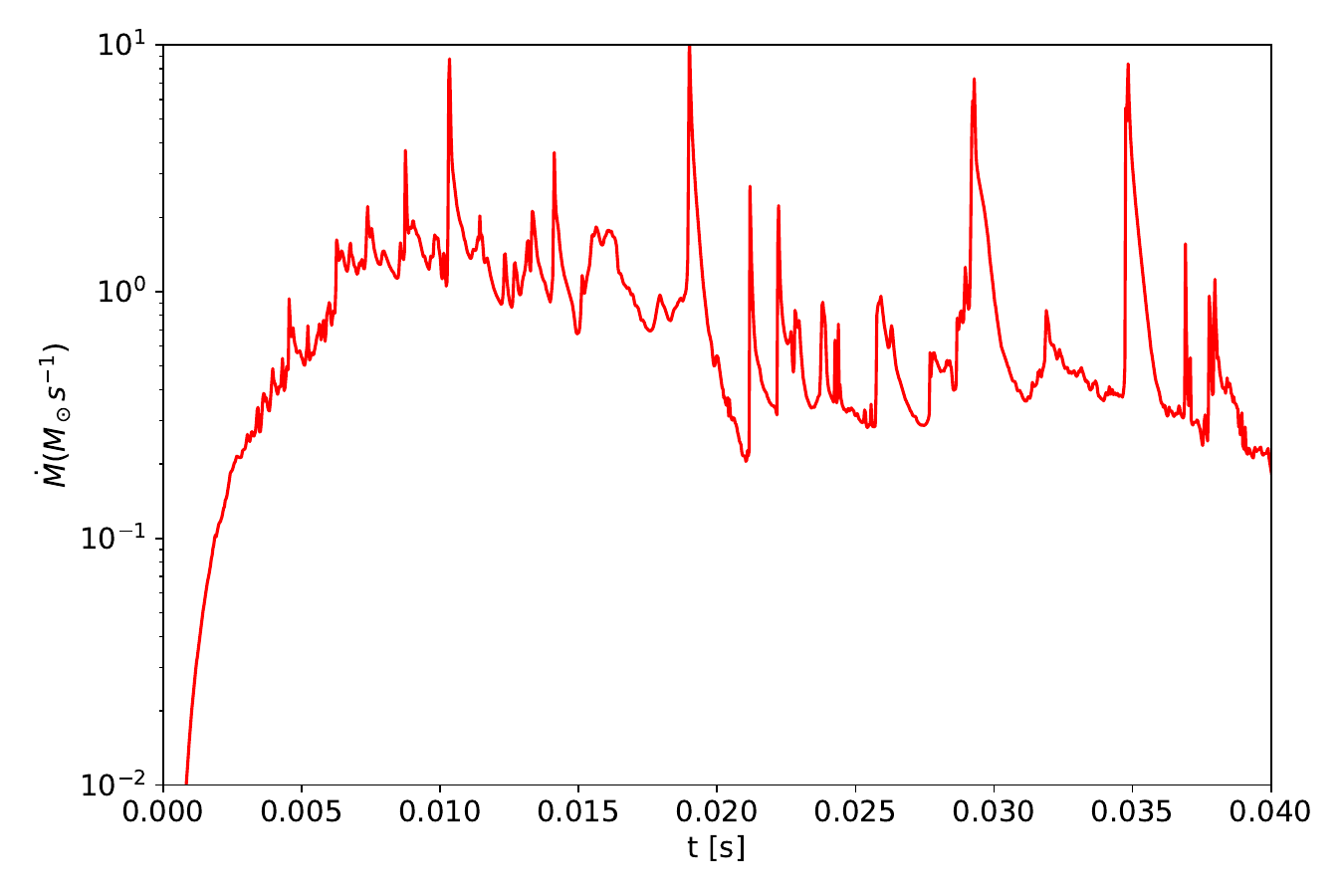}
\caption{The time evolution of the accretion rate for the system with BH mass of $M_{\rm bh}=3M_\odot$ and the initial disk mass of $M_{\rm t}=0.1 M_\odot$. We consider four cases: $a_\bullet=0.1$ (top left), $a_\bullet=0.6$ (top right), 0.9 (bottom left) and 0.98 (bottom right).}\label{fig:AccRate}
\end{figure*}

In Fig. \ref{fig:AccRate}, we plot the accretion rate as the function of time. The instantaneous peaks, which appear in the states of a highly variable accretion rate, reach the values up to several solar masses per second.

%In Fig. \ref{fig:AccRate}, we plot the accretion rate through the BH horizon in the function of time. The instantaneous peaks, which appear in the states of a highly variable accretion rate, reach the values up to several solar masses per second. The flares in the local accretion rate do not necessarily correspond to the observable luminosity flares. In fact, the neutrino luminosity, which we derive by integrating the emissivity over the whole simulation volume, has a smooth dependence on time.

%If the high accretion rate through the black-hole horizon affects the jet production, this rapid variability might have observable consequences. The direct predictions, however, are not possible with the current model.

%The starting mass at $t=0$ was equal to $0.1 M_\odot$.

\section{Neutrino Annihilation Power}
\label{sec:Lvv}
The HARM-COOL code provides the calculation for the total neutrino luminosity, but does not include the estimation for the neutrino annihilation luminosity \citep{Janiuk2017,Janiuk2019}. The later is, however, more important for a direct comparison with the observations.

%% Figure Lvv
\begin{figure*}[ht]
\centering
\includegraphics[width=8cm]{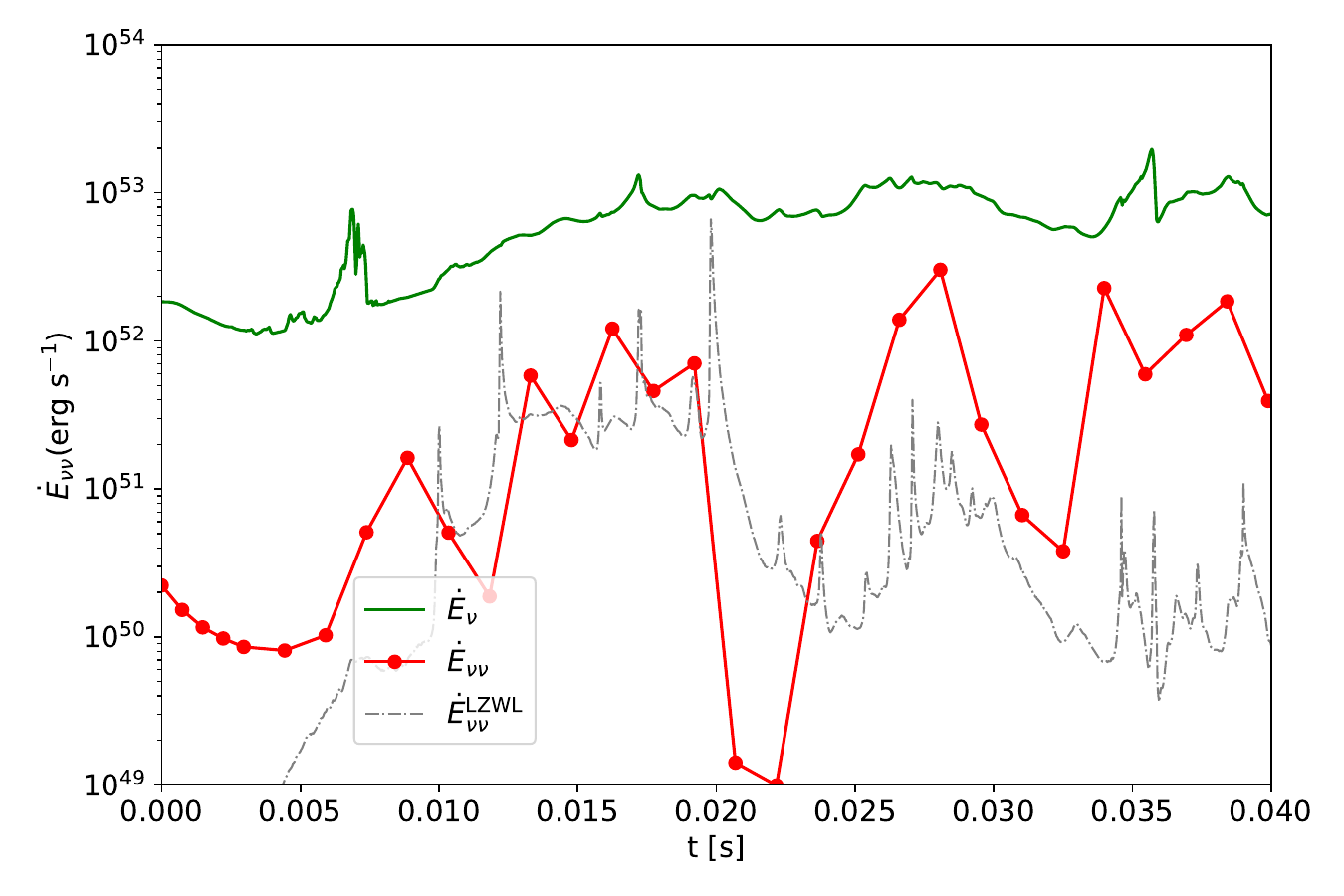}
\includegraphics[width=8cm]{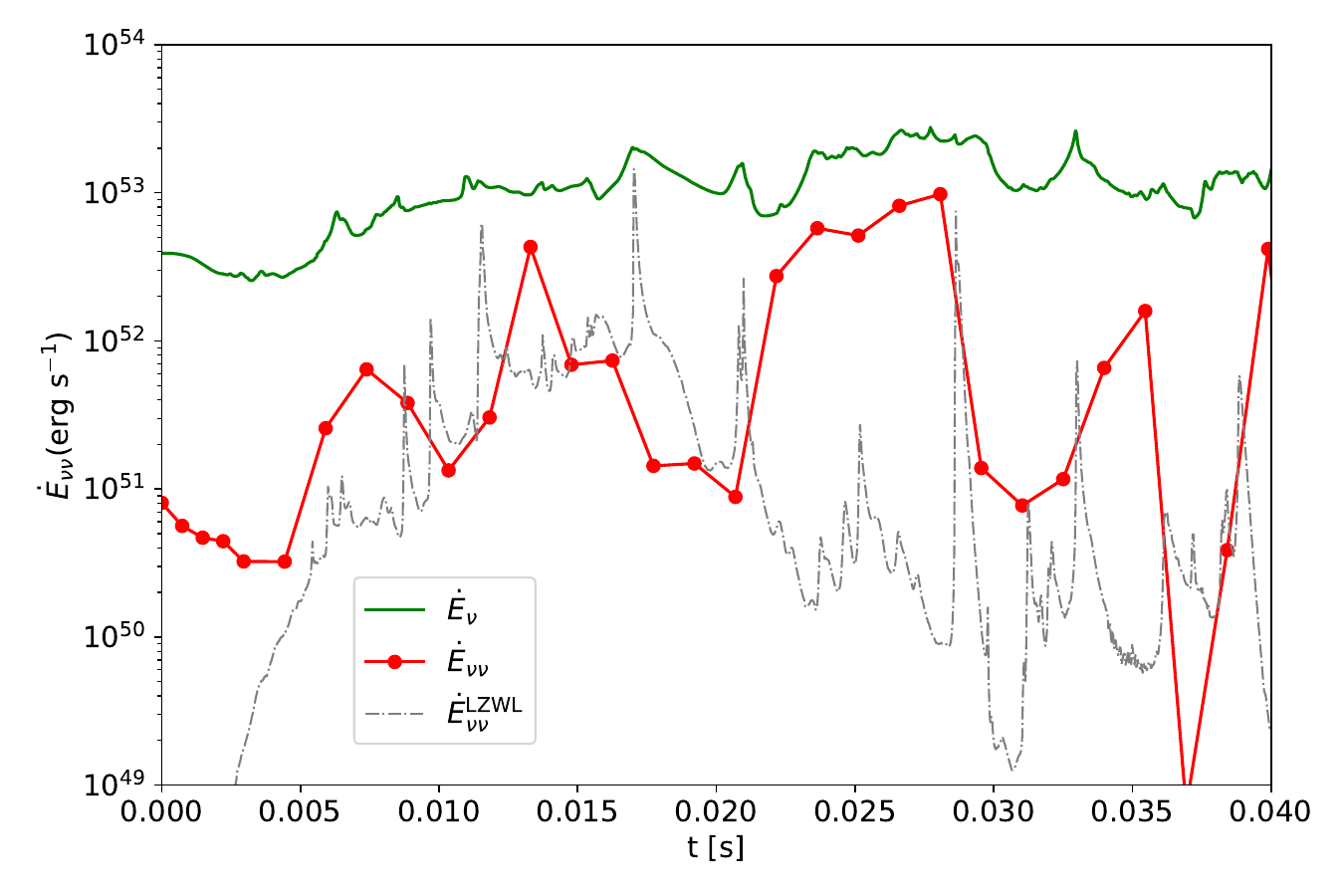}
\includegraphics[width=8cm]{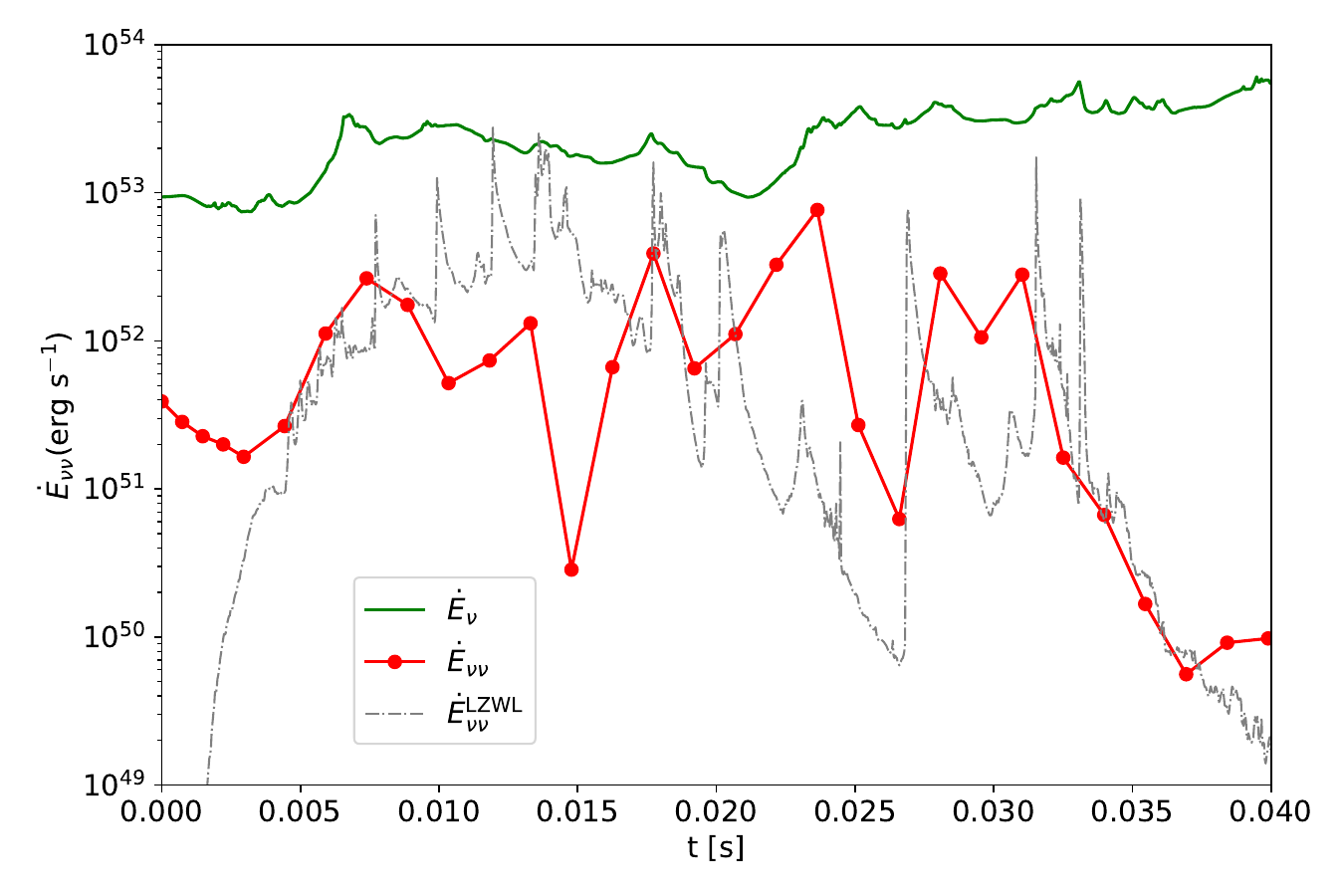}
\includegraphics[width=8cm]{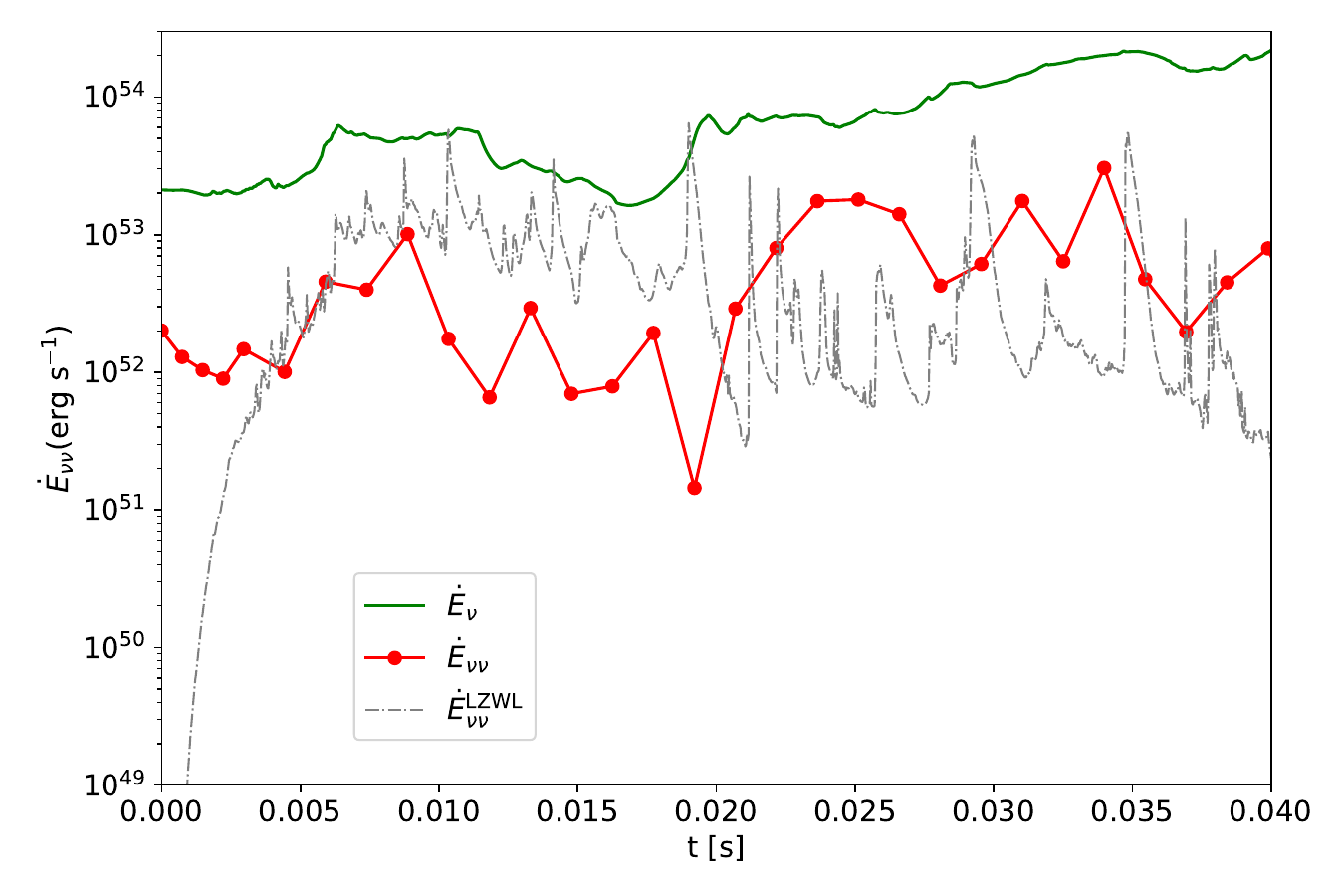}
\caption{The time evolution of the neutrino annihilation power for a BH with mass $m_\bullet=3$, and spin $a_\bullet=0.1$ (top left), $a_\bullet=0.6$ (top right), 0.9 (bottom left) and 0.98 (bottom right).}\label{fig:Lvv}
\end{figure*}

From the HARM-COOL simulations, we can get the neutrino mean energy
$\varepsilon _{\nu _i }^k $ and luminosity $l_{\nu _i }^k $ in each grid cell $k$. The angle at which neutrinos from cell $k$ encounter anti-neutrinos from another cell $k^\prime$ at that point is denoted as $\theta _{k{k^\prime}}$. The height (neutrino annihilation occurs) above (or below) the disk equatorial plane is $d_k $. We write a code to incorporate the neutrino annihilation luminosity. First, the neutrino annihilation lumonosity at a space point is obtained by summating over all pairs of cells \citep{PWF99,Rosswog2003},
\begin{eqnarray}
l_{\nu \bar{\nu}}  = & &  A_1 \sum_k \frac{l^k_{\nu_i}}{d_k^2} \sum_{k^\prime}
\frac{l^{k^\prime}_{\nu_i}}{d_{k^\prime}^2}
(\epsilon^k_{\nu_i}+\epsilon^{k'}_{\bar{\nu}_i})(1-\cos\theta_{kk^\prime})^2
 +   \nonumber \\
& & A_2 \sum_k \frac{l^k_{\nu_i}}{d_k^2} \sum_{k^\prime} \frac{l^{k^\prime}_{\nu_i}}{d_{k^\prime}^2} \frac{\epsilon^k_{\nu_i}+\epsilon^{k\prime}_{\bar{\nu}_i}}{\epsilon^k_{\nu_i} \epsilon^{k^\prime}_{\bar{\nu}_i}}(1-\cos\theta_{kk^\prime}),
\end{eqnarray}
where $A_1 \approx 1.7\times 10^{ - 44} \ {\rm cm \cdot erg^{-2} \cdot s^{-1} }$
and $A_2 \approx 1.6\times 10^{ - 56} \ {\rm cm \cdot erg^{ - 2}s^{-1} }$. The total neutrino annihilation luminosity is given by integrating over the whole space outside the BH and the disk \citep{Liu2018},
\begin{equation}
    \dot{E}_{\nu \bar{\nu}} =4\pi \int\int l_{\nu \bar{\nu}} r dr dz.
\label{eq:Evv}
\end{equation}
% The summation is performed over the whole space outside the BH horizon and the disk. 

The neutrino annihilation power as a function of time is shown in Fig.\ref{fig:Lvv}. We present the results for different BH spins, i.e., $a_\bullet=0.1$ (top left), 0.6 (top right), 0.9 (bottom left) and 0.98 (bottom right), see the solid red lines. We also plot the total neutrino power with solid green lines. Inspecting Fig.\ref{fig:Lvv}, both the neutrino annihilation power $\dot{E}_{\nu \bar{\nu}}$ and the neutrino power $\dot{E}_{\nu}$ on average are scaling with the BH spin.

For comparison, we calculate the neutrino annihilation power from the analytical solutions, i.e., \citep{Lei2017}
\begin{eqnarray}
\dot{E}^{\rm LZWL}_{\nu \bar{\nu}}  \simeq  && \dot{E}_{\nu \bar{\nu}, \rm ign} \left[ \left(\frac{\dot{m}}{\dot{m}_{\rm ign}} \right)^{-\alpha_{\nu \bar{\nu}} } + \left(\frac{\dot{m}}{\dot{m}_{\rm ign} } \right)^{-\beta_{\nu \bar{\nu}} } \right]^{-1} \nonumber \\
&& \times   \left[1+(\frac{\dot{m}}{\dot{m}_{\rm trap} })^{\beta_{\nu \bar{\nu}} - \gamma_{\nu \bar{\nu}}  }\right]^{-1} ,
\label{eq_Evv}
\end{eqnarray}
where,
\begin{eqnarray}
&&\left\lbrace
\begin{tabular}{l}
$\dot{E}_{\nu \bar{\nu}, \rm ign}=10^{(48.0+0.15 a_\bullet)}  \left(\frac{m_\bullet}{3} \right)^{\log(\dot{m}/\dot{m}_{\rm ign}) -3.3} {\rm erg \ s^{-1}},  $ \\
$\alpha_{\nu \bar{\nu}} = 4.7, \ \beta_{\nu \bar{\nu}} =  2.23, \  \gamma_{\nu \bar{\nu}} =0.3, $
\end{tabular} 
\right.\nonumber \\
&& \dot{m}_{\rm ign} = 0.07-0.063 a_\bullet,  \ \dot{m}_{\rm trap} = 6.0-4.0 a_\bullet^3,
\end{eqnarray}
where $m_\bullet=M_\bullet/M_{\odot}$, and  $\dot {m}=\dot{M}/M_{\odot} s^{-1}$. $\dot{m}_{\rm ign}$ and $\dot{m}_{\rm trap}$ are the igniting and trapping accretion rates, respectively. We adopt $m_\bullet=3$, and use the accretion rate $\dot{m}$ from the simulations (see Fig.\ref{fig:AccRate}) for different models, i.e., Mt0.1-a0.1 (top left), Mt0.1-a0.6 (top right), Mt0.1-a0.9 (bottom left) and Mt0.11-a0.98 (bottom right), as shown with dashed gray lines in Fig.\ref{fig:Lvv}. As we can see that the $\dot{E}_{\nu \bar{\nu}}$ obtained based on HARM-COOL is roughly consistent with our previous analytical result \citep{Lei2017}.

\section{Magnetic Power}
\label{sec:LBZ}
Utilizing the magnetic flux $\Phi_{\rm BH}$ through the BH horizon from HARM-COOL simulations, one can estimate the BZ power.

%% Figure Lbz
\begin{figure*}[ht]
\centering
\includegraphics[width=8cm]{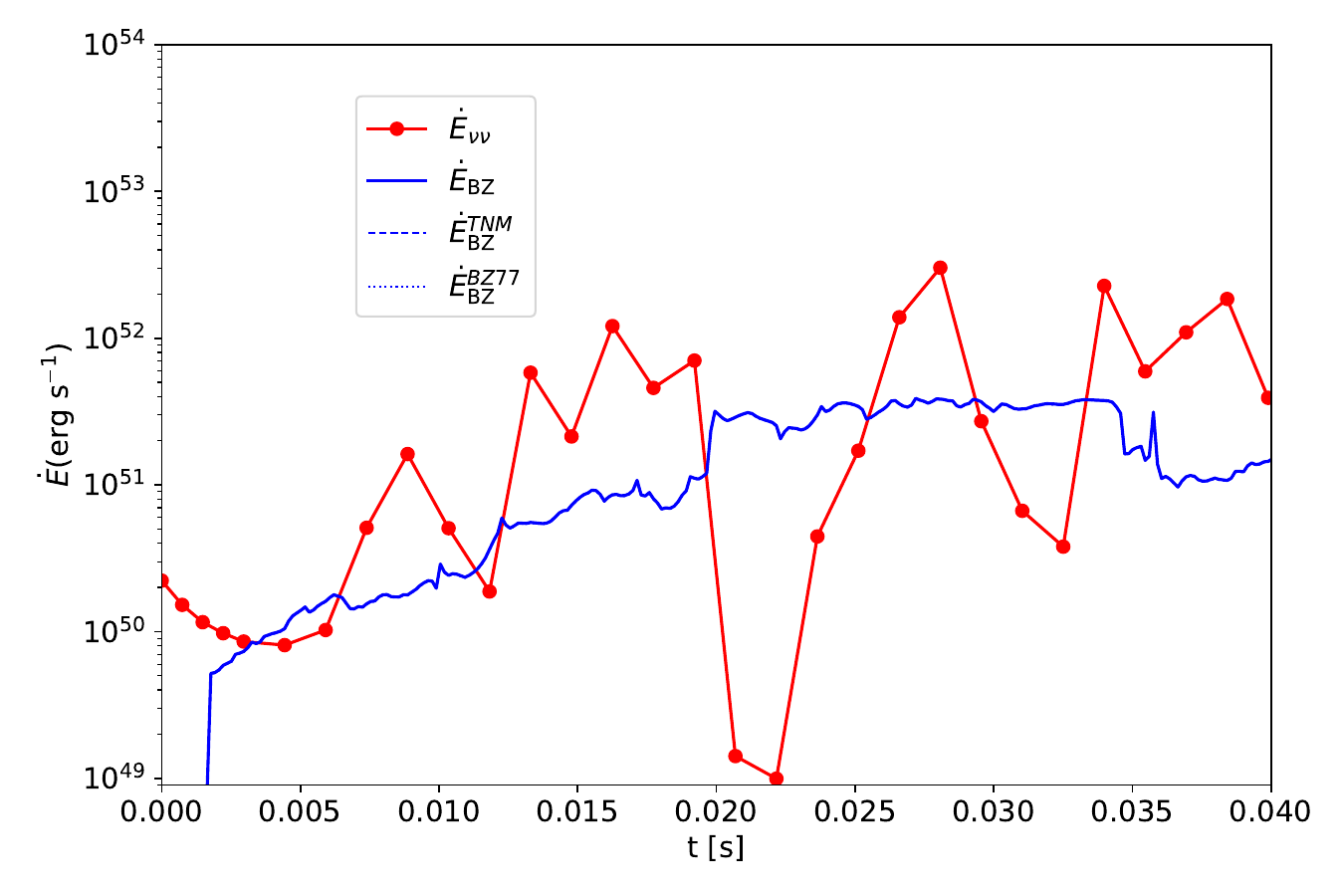}
\includegraphics[width=8cm]{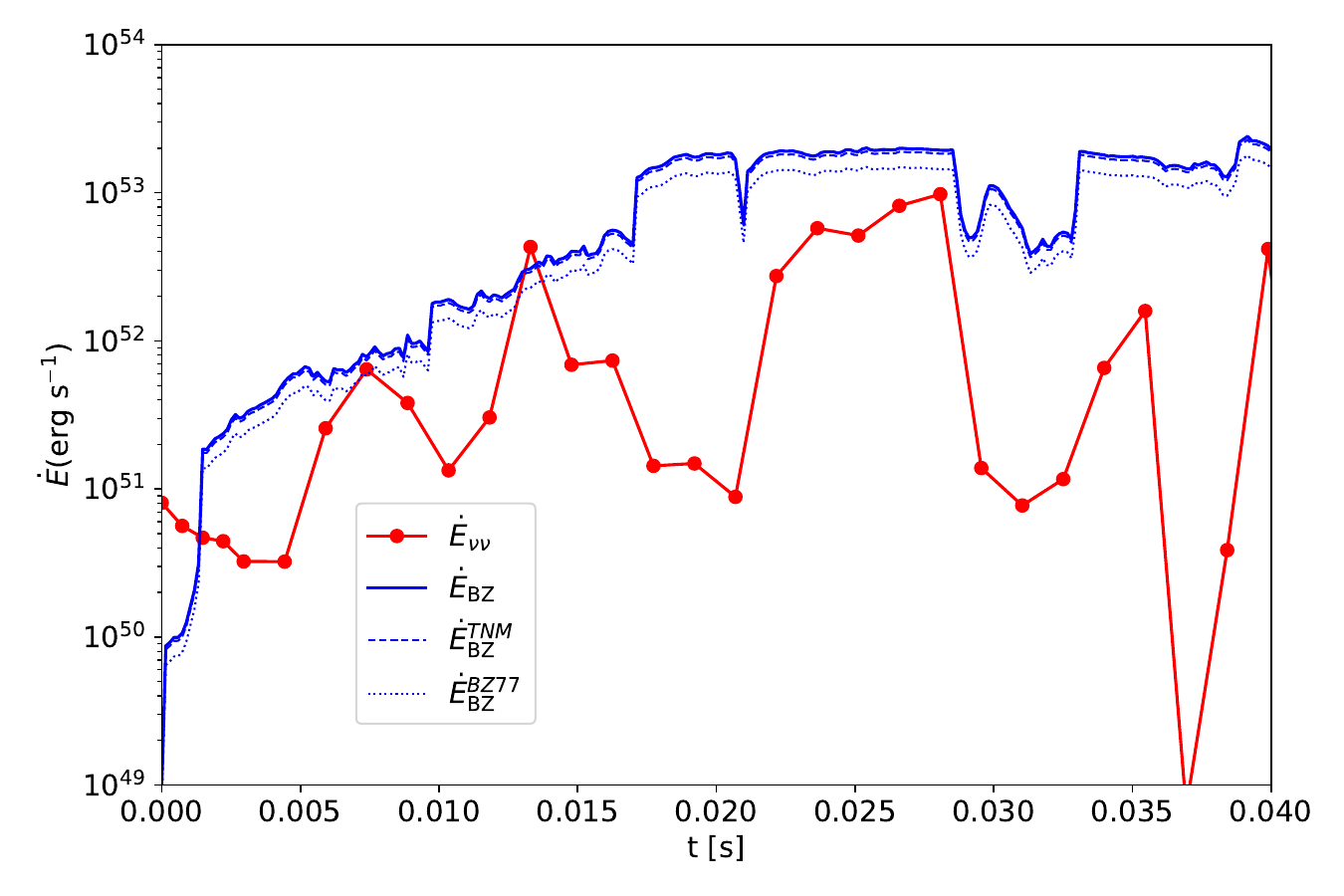}
\includegraphics[width=8cm]{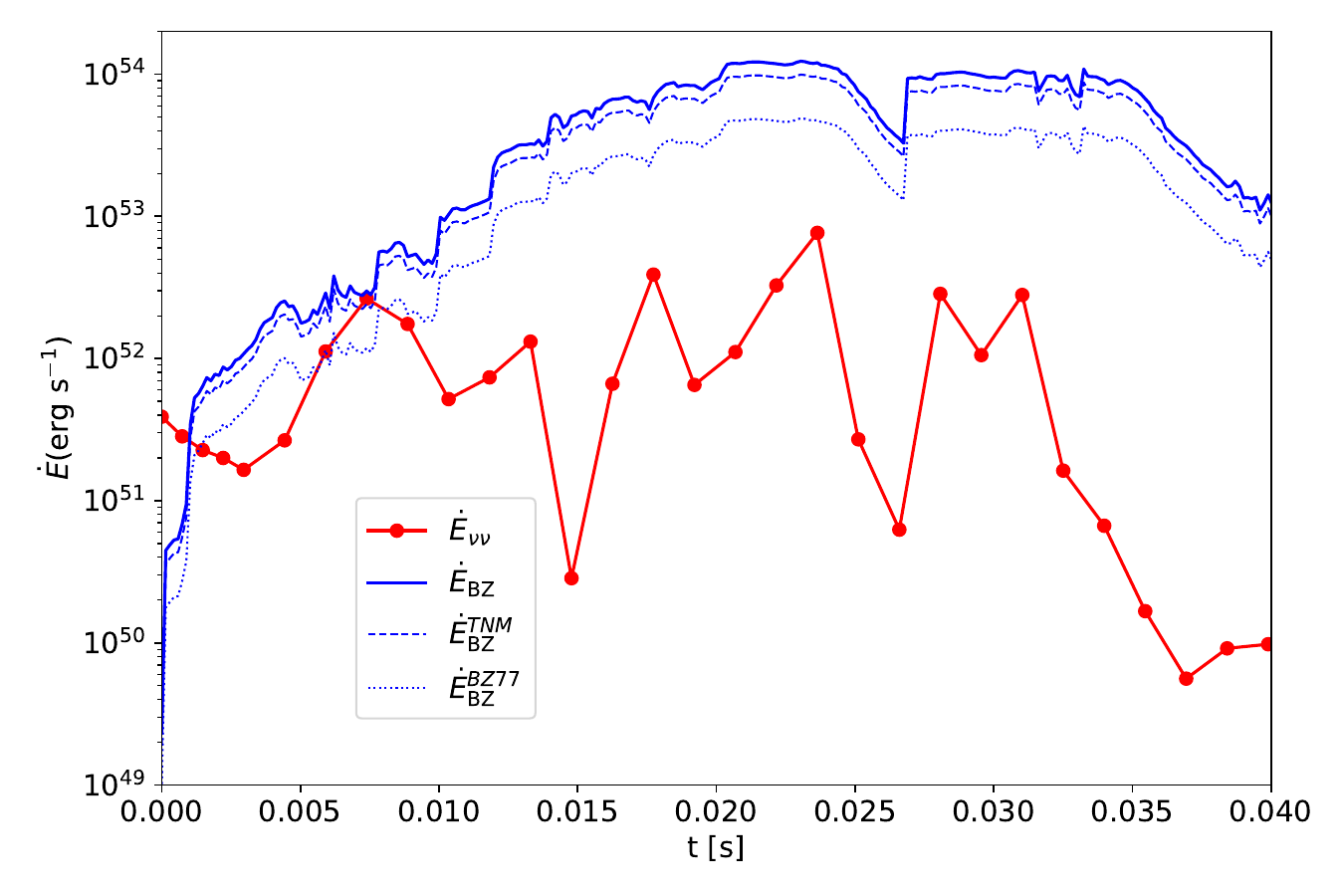}
\includegraphics[width=8cm]{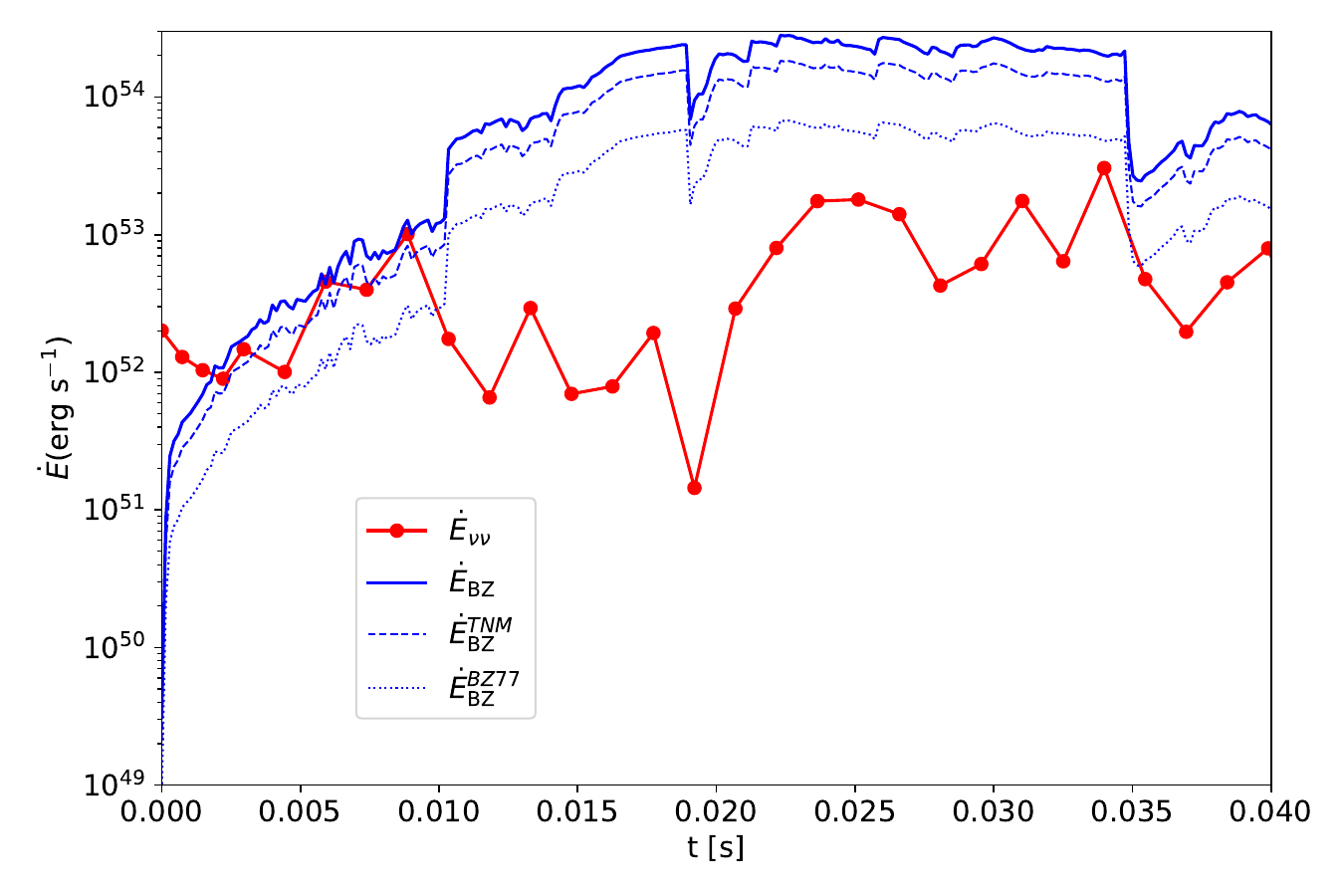}
\caption{The time evolution of the BZ power $\dot{E}_{\rm BZ}$ (solid blue lines) for a BH with mass $m_\bullet=3$, and spin $a_\bullet=0.1$ (top left), $a_\bullet=0.6$ (top right), 0.9 (bottom left) and 0.98 (bottom right). For comparisons, we also show the results of $\dot{E}_{\rm BZ}^{\rm BZ77}$ (dotted blue lines), $\dot{E}^{\rm TNM}_{\rm BZ}$ (dashed blue lines), and the neutrino power $\dot{E}_{\nu \bar{\nu}}$ (red lines).  }\label{fig:Lbz}
\end{figure*}

The BZ jet power from a BH with mass $M_{\bullet}$ and angular momentum $J_\bullet$ is \citep{Lee2000,Li2000,Wang2002,Lei2005,Lei2011,Lei2013,Liu2015}
\begin{equation}
\dot{E}_{\rm BZ}= \frac{c}{4\pi} \Phi_{\rm BH}^2 \frac{q^2     F(a_\bullet)}{16 \pi r_{\rm g}^2}, 
\label{eq:BZ}
\end{equation}
where $F(a_\bullet)=[(1+q^2)/q^2][(q+1/q) \arctan q-1]$, $q= a_{\bullet} /(1+\sqrt{1-a^2_{\bullet}})$, and $\Phi_{\rm BH}$ is an absolute magnetic flux through the BH. Here, $2/3\le F(a_\bullet)\le \pi-2$ for $0\le a_\bullet \le 1$. The evolution of BZ power for different models is plotted in Fig.\ref{fig:Lbz} with solid blue lines.

For comparison, we also plot the other expressions.
\citet{BZ77} derived the magnetic power of a force-free jet for a slowly spinning BH ($a_\bullet \ll 1$) \citep{BZ77,Barkov2011}
\begin{equation}
 \dot{E}_{\rm BZ}^{\rm BZ77} = \frac{\kappa c}{4\pi} \Phi_{\rm BH}^2 \frac{a_\bullet^2}{16 r_{\rm g}^2} ,
 \label{eq:BZ77}
\end{equation} 
where $\kappa$ depends weakly on the field geometry (it is 0.053 for a split monopole geometry and 0.044 for a parabolic geometry). $\kappa=0.053$ is adopted in the calculations. We find that Eq.(\ref{eq:BZ}) will become Eq.(\ref{eq:BZ77}) when $a_\bullet \ll 1$. The results for $\dot{E}_{\rm BZ}^{\rm BZ77}$ are shown with dotted blue lines in Fig.\ref{fig:Lbz}.

Based on GRMHD simulations, \citet{McKinney2005} showed that for $a_\bullet \ge 0.5$ the BZ power varies as steeply as the fourth power of the BH angular velocity $\Omega_\bullet$, i.e., $\dot{E}_{\rm BZ} \propto \Omega_\bullet^4$, where $\Omega_\bullet=a_\bullet c/(2r_\bullet)$ and the radius of the BH horizon $r_\bullet=(1+\sqrt{1-a_\bullet^2})r_{\rm g}$. In addition, \citet{TNM2010} further explored this scaling in detail using GRMHD simulations, and found an even steep scaling $\dot{E}_{\rm BZ} \propto \Omega_\bullet^6$ for thicker disks. 
\citet{TNM2010}, therefore, extended the magnetic power in \citet{BZ77} to high-spin BHs and obtained \citep{McKinney2005, TNM2010,Tchekhovskoy2012}
\begin{equation}
\dot{E}^{\rm TNM}_{\rm BZ} = \frac{\kappa}{4\pi c} \Phi_{\rm BH}^2 \Omega_\bullet^2 f(\Omega_\bullet), 
\label{eq:BZ_TNM}
\end{equation}
where $f(\Omega_\bullet) \simeq 1 + 1.38 (\Omega_\bullet r_{\rm g}/c)^2 - 9.2 (\Omega_\bullet r_{\rm g}/c)^4$ is a high-spin correction to Eq.(\ref{eq:BZ77}). For $a_\bullet \ll 1$, we find that $\Omega_\bullet \sim a_\bullet c/(4 r_{\rm g})$ and $f(\Omega_\bullet)\sim 1$, and Eq.(\ref{eq:BZ_TNM}) returns to Eq. (\ref{eq:BZ77}). The results for $\dot{E}_{\rm BZ}^{\rm TNM}$ are shown with dashed blue lines in Fig.\ref{fig:Lbz}.

We present the results for different BH spins, i.e., $a_\bullet=0.1$ (top left), $a_\bullet=0.6$ (top right), 0.9 (bottom left) and 0.98 (bottom right). The power of BZ $\dot{E}_{\rm BZ}$ with the formula of Eq. (\ref{eq:BZ}) is found to be quite close to $\dot{E}_{\rm BZ}^{\rm TNM}$ in all cases. However, the expression $\dot{E}_{\rm BZ}^{\rm BZ77}$ can only apply to the case with low BH spin. In high BH spin case (like $a_\bullet=0.98$), $\dot{E}_{\rm BZ}^{\rm BZ77}$ is significantly smaller than $\dot{E}_{\rm BZ}$. For $a_\bullet=0.1$, the difference between their values ($\dot{E}_{\rm BZ}$, $\dot{E}_{\rm BZ}^{\rm BZ77}$ and $\dot{E}_{\rm BZ}^{\rm TNM}$) is nearly indistinguishable.

For comparison, we also show the evolution of the neutrino annihilation power $\dot{E}_{\nu \bar{\nu}}$ in Fig. \ref{fig:Lbz} (see the red lines). The neutrino annihilation power is always dominated at the beginning, since the magnetic flux is still building. The BZ luminosity gradually becomes dominant during the evolution for high spin BH. For $a_\bullet=0.1$, the jet luminosity is dominated by $\dot{E}_{\nu \bar{\nu}}$ for some time intervals, while by $\dot{E}_{\rm BZ}$ in other episodes, see the top left panel of Fig. \ref{fig:Lbz}.

%% Figure sigma
\begin{figure}[ht]
\centering
\includegraphics[width=9cm]{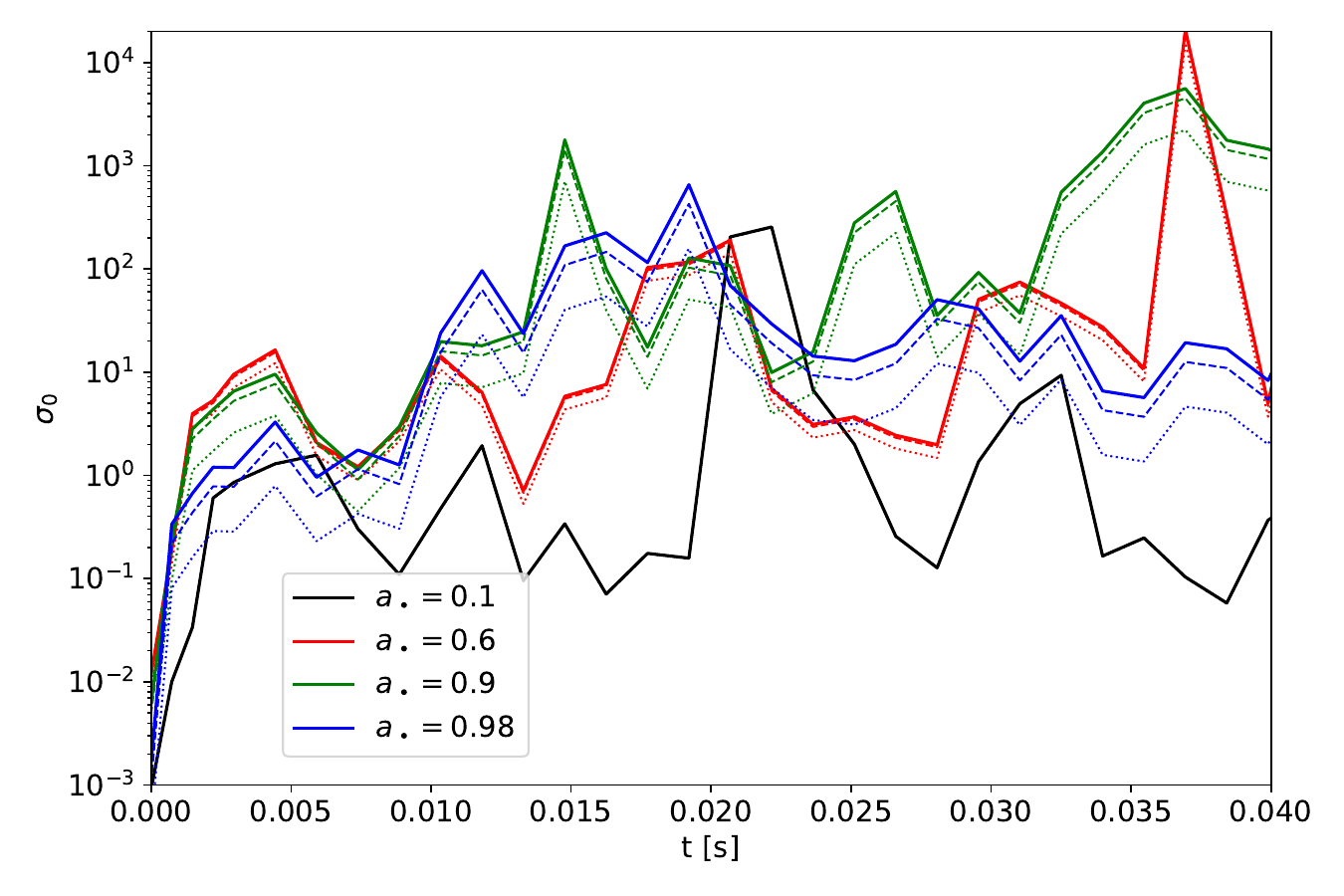}
\caption{The time evolution of the magnetization parameter $\sigma_0$ for a BH with mass $m_\bullet=3$, and spin $a_\bullet=0.1$ (black), $a_\bullet=0.6$ (red), $a_\bullet=0.9$ (green), and $a_\bullet=0.98$ (blue). The solid, dotted and dashed lines represent $\sigma_0$ calculated with $\dot{E}_{\rm BZ}$, $\dot{E}_{\rm BZ}^{\rm BZ77}$ and $\dot{E}^{\rm TNM}_{\rm BZ}$, respectively. }\label{fig:sigma}
\end{figure}

We can define the magnetization parameter,
\begin{equation}
    \sigma_0 = \frac{\dot{E}_{\rm BZ}}{\dot{E}_{\rm m}},
\end{equation}
where $\dot{E}_{\rm m} = \dot{E}_{\nu \bar{\nu}}+ \dot{M}_{\rm j} c^2$ is the total matter energy outflow luminosity, and $\dot{M}_{\rm j}$ is the baryon loading rate for GRB jet \citep{Lei2013,Lei2017}. In GRB prompt emission phase, $\dot{E}_{\nu \bar{\nu}}$ is usually much larger than $\dot{M}_{\rm j} c^2$, leading to an extreme relativistic jet. Therefore, we take $\sigma_0 \sim \dot{E}_{\rm BZ}/\dot{E}_{\nu\nu}$. The time evolution of $\sigma_0$ is presented in Fig.\ref{fig:sigma} for Mt0.1-a0.1 (black lines), Mt0.1-a0.6 (red lines), Mt0.1-a0.9 (green lines) and Mt0.11-a0.98 (blue lines) models. For comparison, $\sigma_0$ calculated with $\dot{E}_{\rm BZ}$, $\dot{E}_{\rm BZ}^{\rm BZ77}$ and $\dot{E}^{\rm TNM}_{\rm BZ}$ are also presented with solid, dotted, and dashed lines, respectively. We find a highly variable $\sigma_0$. On average, $\sigma_0 \sim 1$ for $a_\bullet=0.1$, while $\sigma_0 > 1$ for high spin cases.

The baryon loading is another fundamental parameter of GRB, I plan to address this point in future work. Once the baryon-loading rate $\dot{M}_{\rm j}$ is given, we can then define a parameter denoting the maximum available energy per baryon in the jet driven by the BZ process, 
\begin{equation}
\mu_0 \equiv \frac{\dot{E}}{\dot{M}_{\rm j} c^2} = \frac{\dot{E}_{\rm m}+\dot{E}_{\rm BZ}}{\dot{M}_{\rm j} c^2} = \eta (1+\sigma_0),
\label{eq:mu}
\end{equation}
where $\eta \equiv \frac{\dot{E}_{\rm m}}{ \dot{M}_{\rm j} c^2 }$ is the dimensionless ``entropy'' parameter. The acceleration behavior of the jet is subject to uncertainties. Generally, the jet will reach a terminating Lorentz factor $\Gamma$ that satisfies $\Gamma_{\rm min} < \Gamma < \Gamma_{\rm max}$ with the explicit value depending on the detailed dissipation process, such as kink instability (Wang et al. 2006) and ICMART \citep{Zhang2011}. Here, $\Gamma_{\rm min}=\max(\mu_{0}^{1/3},\eta)$ and $\Gamma_{\rm max} =  \mu_0$, which correspond to the beginning and the end of the slow acceleration phase in a hybrid outflow, respectively \citep{Gao2015}.

\section{Discussions}
\label{sec:discussion}
The central engine of GRBs is likely a hyperaccreting BH. The neutrino annihilation and BZ processes are two candidate mechanisms for powering GRB jets. In this paper, we extended the HARM-COOL code to include the calculation of neutrino annihilation power, and studied the time evolution of the central engine parameters for these two models.

%performed 2D GRMHD simulations for the hyperaccreting BH with HARM-COOL, and 

%obtained the analytical solutions to the neutrino and magnetic models, and studied the time evolution of the central engine parameters for these two models.

The evolution of $\dot{E}_{\nu \bar{\nu}}$, $\dot{E}_{\rm BZ}$ and $\sigma_0$ exhibits high variability. The $\dot{E}_{\nu \bar{\nu}}$ obtained based on HARM-COOL is roughly consistent with our previous analytical result \citep{Lei2017}. We also find that the neutrino-annihilation power $\dot{E}_{\nu \bar{\nu}}$ increases with BH spin. The BZ power is calculated by using the magnetic flux from the code and with three different analytical expressions. The difference of $\dot{E}_{\rm BZ}$ with the different expressions is significant for high spin BH and can be ignored for very low spin case.

\begin{figure*}[ht]
\centering
\includegraphics[width=8cm]{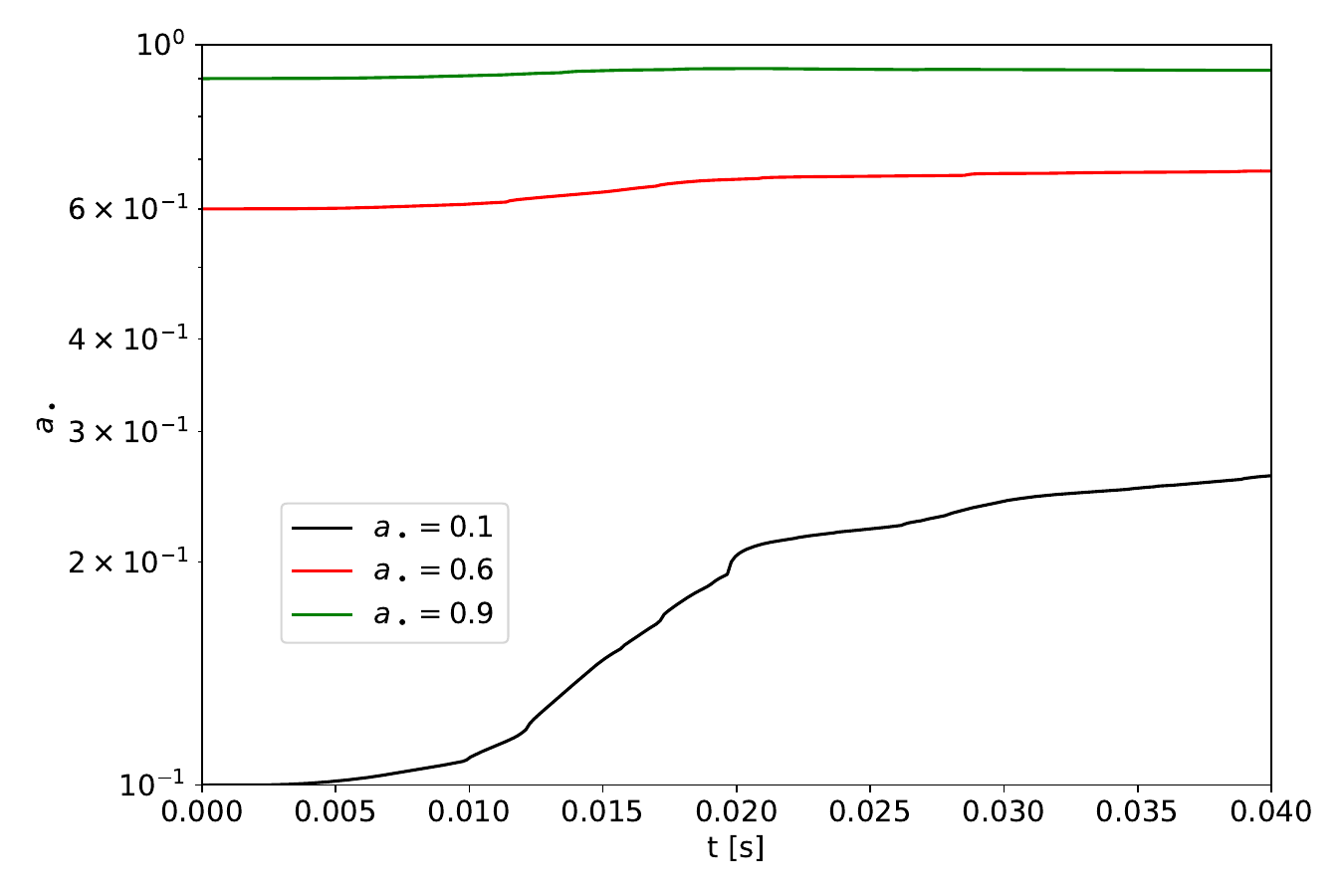}
\includegraphics[width=8cm]{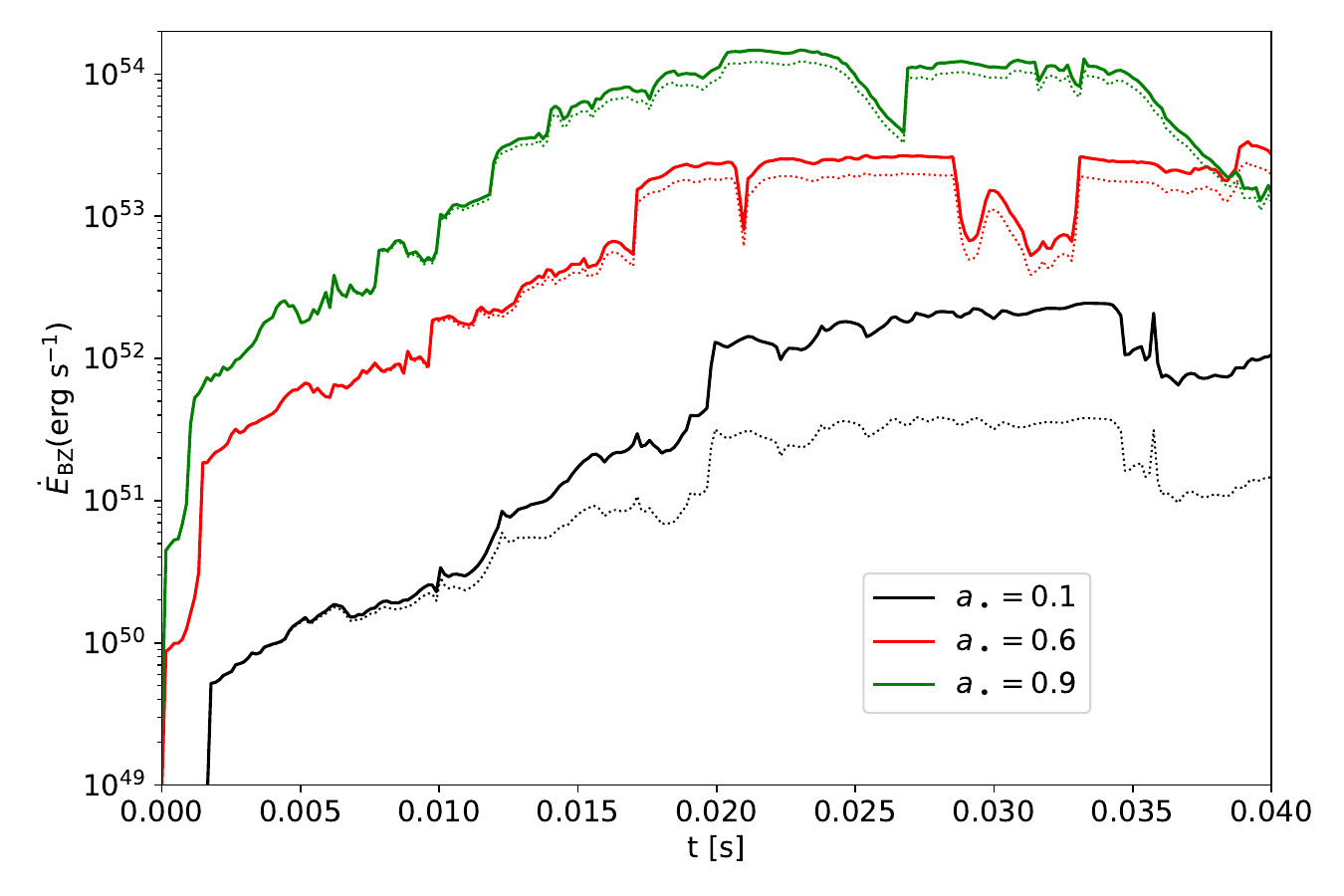}
\caption{The time evolution of the BH spin $a_\bullet$ (left) and the BZ power $\dot{E}_{\rm BZ}$ (right) for a BH with initial mass $m_\bullet=3$, and initial spin $a_\bullet=0.1$ (black), $a_\bullet=0.6$ (red), and $a_\bullet=0.9$ (green). For comparison, we also show the evolution of the BZ power with constant mass and spin with dotted lines. }\label{fig:aev}
\end{figure*}

In this paper, we conduct the simulation in a fixed background metric. Instead of using a stationary metric with the BH having constant mass and spin and simply allowing the matter to accrete onto it, \citet{Janiuk2018} extend the code to account for a changing Kerr metric. This will be included in future work. For a first step, we can investigate the evolution of BH in a simple way based on the current simulation results. Considering both the accretion and the BZ processes, the evolution equations are given by \citep{Wang2002} 
\begin{equation}
\frac{dM_\bullet c^2}{dt} = \dot{M} c^2 E_{\rm ms} - \dot{E}_{\rm BZ},
\label{dMbz}
\end{equation}

%\begin{equation}
%\frac{dJ_\bullet}{dt} = \dot{M} L_{\rm ms} - T_{\rm BZ}
%\label{dJbz}
%\end{equation}
%the evolution equation for the BH spin is then
\begin{eqnarray}
\frac{da_\bullet}{dt} = && (\dot{M} L_{\rm ms} - T_{\rm BZ})c/(G M_\bullet^2) - \nonumber \\
&& 2 a_\bullet (\dot{M} c^2  E_{\rm ms} - \dot{E}_{\rm BZ}) /(M_\bullet c^2),
\end{eqnarray}
where $T_{\rm BZ} = \frac{\dot{E}_{\rm BZ}}{\Omega_{\rm F}}$ is the BZ torque applied on the BH, and $\Omega_{\rm F}=0.5\Omega_\bullet$ is usually taken to maximize the BZ power. Here, $E_{\rm ms}$ and $L_{\rm ms}$ are the specific energy and the specific momentum corresponding to the inner most radius $r_{\rm ms}$ of the disk \citep{Bardeen1972}, which are defined as
$E_{\rm ms} = (4\sqrt{ R_{\rm ms} }-3a_{\bullet}) /(\sqrt{3} R_{\rm ms})$, $L_{\rm ms} = (G M_\bullet/c) (2 (3 \sqrt{R_{\rm ms}} -2 a_\bullet) )/(\sqrt{3} \sqrt{R_{\rm ms}} )$, where $R_{\rm ms} = r_{\rm ms}/r_{\rm g} $. In our simulations, we adopt a torus with only a mass of $0.1 M_\odot$, the BH mass $M_\bullet$ will therefore not change significantly. Here, we just focus on the evolution of BH spin and its effects on BZ power, as shown in Fig.\ref{fig:aev} (solid lines). For comparison, we also show the evolution of the BZ power with constant mass and spin, see the dotted lines. Inspecting Fig.\ref{fig:aev}, we can find significant increases in both the BH spin and the BZ power for the low initial spin case.

%$\Omega_\bullet =\frac{a_\bullet c}{2 r_\bullet}= \frac{c^3}{G M_\bullet} \frac{a_\bullet}{2 (1+\sqrt{1-a_\bullet^2})}$ is the angular velocity of BH horizon.

%% Figure evolution

%The total magnetic torque applied on the BH is (Lee et al. 2000; Li 2000; Wang et al. 2002; McKinney 2005; Lei et al. 2005; Lei \& Zhang 2011; Lei et al. 2013)
%\begin{eqnarray}
%T_{\rm B} = \frac{\dot{E}_{\rm B}}{\Omega_{\rm F}}=3.4 \times 10^{45} a_\bullet^2 q^{-1} m_\bullet^3 B_{\bullet,15}^2  F(a_\bullet) {\rm \ g \ cm^2 \ s^{-2}}, \nonumber \\
%\end{eqnarray}
%where $\Omega_{\rm F}=0.5\Omega_\bullet$ is usually taken to maximize the BZ power, and
%\begin{equation}
%\Omega_\bullet =\frac{a_\bullet c}{2 r_\bullet}= \frac{c^3}{G M_\bullet} \frac{a_\bullet}{2 (1+\sqrt{1-a_\bullet^2})}
%\end{equation}
%is the angular velocity of BH horizon.

%As a BH may be spun up by accretion or spun down by the BZ mechanism, the BH spin will reach an equilibrium value when $da_\bullet/dt =0$. If the magnetic field is related to the mass accretion rate as Equation (\ref{Bmdot}), the final BH spin will be $a_\bullet^{\rm eq} \sim 0.87$.

%The evolution of BH spin combining with the accretion profile will give rise to a reasonable GRB lightcurve. In addition, possible jet pression (Lei et al. 2007), episodic jet (Yuan \& Zhang 2012) and episodic accretion (by magnetic barrier, see Proga \& Zhang 2006; or by magnetically arrested disk (MAD), see Lloyd-Ronning et al. 2016) would enrich the structure of lightcurve.

As discussed in \citet{Lei2013,Lei2017}, the neutrino annihilation mechanism launches a thermal ``fireball'', leaving an imprint of the thermal component in the GRB spectrum. However, a jet powered by the BZ mechanism is Poynting-flux-dominated. In this paper, we find that the neutrino annihilation power is dominated at the beginning, and the BZ luminosity gradually becomes dominant during the evolution for high spin BH. For very low spin case, like $a_\bullet=0.1$, the jet luminosity is dominated by $\dot{E}_{\nu \bar{\nu}}$ for some time intervals, while by $\dot{E}_{\rm BZ}$ in other episodes. The prompt emission of GRB 200613A show a significant spectral evolution from a fireball to a Pyontying-flux-dominated jet \citep{ZhangBB2018NatAs}, which might be interpreted with the building-up of magnetic flux or spin-up of the central BH \citep{Liu2012,Gao2022}. Recently, \citet{Fu2024} performed a time-resolved spectral analysis of the prompt emission of GRB 200613A, and revealed that the BB components are only significant in some time intervals and do not appear in other time periods. Such evolution behavior of prompt spectra could be understood with our simulation results by relating the BB component to the neutrino annihilation power and the Band component to the BZ power \citep{Liu2015,Li2024}. As to $\sigma_0$, \citet{Gao2015} also found the variation of the magnetic parameter during the prompt phase of GRB 110721A. 

In this paper, we ignore the baryon loading, which is also a fundamental parameter of GRB. In our in previous works, we derived the baryon loading rate and the evolution based on analytical solutions \citep{Lei2013,Lei2017}.  \citet{Sapountzis2019} pointed out that the formation of a magnetic barrier leads to the launch of a low baryon jet and estimated the maximum achievable Lorentz factor in the jets produced by simulations \citep{Janiuk2021}. However, they mainly focused on the purely electromagnetic effects and ignored the potential neutrino implications on the launching process. Our estimation of neutrino annihilation power will remain a good extension for this problem.

%For very low spin case, like $a_\bullet=0.1$, the jet luminosity is dominated by $\dot{E}_{\nu \bar{\nu}}$ for some time intervals, while by $\dot{E}_{\rm B}$ in other episodes, see the top left panel of Fig. \ref{fig:Lbz}.

%The prompt emission of GRB 200613A show a significant spectral evolution. The spectrum can be best described with Band+BB model. The BB components are only significant in some time intervals, and do not appear in other time periods \citep{Fu2024}. Such evolution behavior of prompt spectra could be understood with our simulation results by relating the BB component to the neutrino annihilation power and the Band component to the BZ power. 

%For a BH central engine with a small initial spin a•(0), the jet might be first dominated by the neutrino-annihilation power and then by the BZ power, leading to a transition from a thermally dominated fireball to a Poynting-flux-dominated flow, as is observed in some GRBs, e.g., GRB 160625B.

%There are several predictions in our model, such as the transition from a thermal- to a magnetic-dominated jet, the evolution of μ0, and the late-time plateaus. Systematic comparisons of these predictions against a large GRB sample are needed to test the BH central engine models. Some examples (e.g., GRB 160625B and GRB 110721A) that are consistent with our model predictions have been observed.

GRBs are also expected to occur in dense environments, e.g., the accretion disk of active galactic nuclei (AGN) \citep{Zhu2021,Perna2021, Liu2021,YuanC2022,Yuan2025}. \citet{Kathirgamaraju2024} performed GRMHD simulations (with HARM) of a binary neutron star mergers occurring within AGN disks, and studied the jet launch and observability of GRB afterglow. We can also include the neutrino process in these studies.

%In this paper, we ignore the baryon loading during the late time central engine activities, since there is no good knowledge on the thermally driven wind at low accretion rates when neutrino cooling totally shuts off. Our analytical solutions are based on the numerical results of a standard NDAF model. We did not include the effects, such as magnetic coupling (Lei et al. 2009), inner boundary torque (Xie et al. 2016) and vertical structure (Liu et al. 2014). These effects may be important, but usually depend on some uncertain parameters. GRMHD simulations will help to give a better understanding of these issues.

%\appendix
%\noindent\\
%\textbf{Acknowledgments}\\
\section*{Acknowledgements}
I thank the anonymous referee for helpful suggestions. I am very grateful to Agnieszka Janiuk, He Gao and Bing Zhang for their helpful discussions. This work is supported by the National Natural Science Foundation of China under grants 12473012 and 12533005, and the National Key R\&D Program of China (No. SQ2023YFC220007).
%\end{acknowledgments}

%% If you have bib database file and want bibtex to generate the
%% bibitems, please use
%%
\bibliographystyle{elsarticle-harv}
\bibliography{refs}

%% else use the following coding to input the bibitems directly in the
%% TeX file.

%% Refer following link for more details about bibliography and citations.
%% https://en.wikibooks.org/wiki/LaTeX/Bibliography_Management

%\begin{thebibliography}{00}

%% For authoryear reference style
%% \bibitem[Author(year)]{label}
%% Text of bibliographic item

%\bibitem[Lamport(1994)]{lamport94}
%  Leslie Lamport,
%  \textit{\LaTeX: a document preparation system},
%  Addison Wesley, Massachusetts,
%  2nd edition,
%  1994.

%\end{thebibliography}
\end{document}